\documentclass[aps,prb,floatfix,twocolumn,showpacs]{revtex4}
\usepackage{latexsym}
\usepackage{graphicx}
\newcommand{\figwidth}{3.275 in}
\usepackage{subfigure}    
\usepackage{epsfig}
\begin{document}
\title{Finite temperature spectral function of a hole in a quantum 
antiferromagnet and role of phonons} 
\author{Satyaki Kar$^{(1)}$ and Efstratios Manousakis$^{(1,2)}$ }
\affiliation{$^{1}$Department of Physics and MARTECH, 
Florida State University, 
Tallahassee, FL 32306-4350, USA and \\
$^{2}$Department of  Physics, University of Athens,
Penipistimiopolis, Zografos, 157 84 Athens, Greece.}  
\date{\today}
\begin{abstract}
We study thermal broadening of  the hole spectral function of the 
two-dimensional $t-J$  model (and its extensions) 
within the non-crossing approximation with and 
without the contribution of optical phonons. 
We  find that phonons at finite
temperature broaden  the lowest energy quasiparticle peak, however,
the string excitations survive  even for relatively strong
electron-phonon coupling. Experimental angle resolved photo-emission
spectroscopy(ARPES)  results  compare well  with  our calculations  at
finite temperature when we use  strong electron-phonon coupling without
any adhoc broadening. 
In addition, we have studied the role of vertex corrections and we
find that their contribution allows us achieve the same overall 
agreement with the ARPES experimental results but using smaller
values for the electron-phonon coupling. 
\end{abstract}
\maketitle

\section{Introduction}

%The  continuous   transition  of  antiferromagnetic   insulator  to  a
%superconductor upon hole/electron doping,  which is the common feature
%of the high temperature  superconducting(HTS) compounds has been a hot
%topic of research in the condensed matter community for more more than
%a couple  of decades  now. As  2D $CuO_2$ planes  in HTS  compunds are
%found to have the most interesting physics lying in them, studies have
%been  done using 2D  Hubbard model,  t-J models  etc. starting  with a
%single  hole  in  a  system   in  order  to  understand  the  role  of
%doping. 
The cuprous oxide superconductors show a broad peak
near the Fermi energy  followed by a ``waterfall''-like  feature at higher
energies in  rather recent high resolution
angle-resolved-photoemission-spectroscopy(ARPES)  
measurements\cite{damascelli,Ronning,Graf}.  Calculations based on the 
  $t-J$ model  give  a well-defined  quasiparticle-like  low energy 
peak and higher energy ``string-like'' excitations\cite{liu,strings}. 
The results obtained from the $t-J$ and the $t-t'-t''-J$ models,
using an artificial broadening of the lowest energy peak
and  of the other peaks corresponding to the string-excitations,
agree reasonably well with  the experimental  spectra\cite{strings}.
Furthermore, there are similar studies using the Hubbard model
and its extensions\cite{Srivastava,macridin,zemljic} 
also indicating that the above
features seen in the ARPES studies could be due to higher
energy hole excitations arising naturally in these strongly correlated
electronic models. 

In this paper we consider the role of finite temperature and of 
the coupling of the hole to optical phonons, 
 as recent experiments have provided increasing evidence that
electron-phonon coupling is strong in 
cuprates\cite{damascelli,Lanzara,Kim,Zhang}.
Our goal is to examine (a) whether or not the string excitations,
claimed in Ref.~\onlinecite{strings} to be the cause of the ``waterfall''-like 
features seen in  the ARPES studies, 
survive the presence of such strongly coupled phonons
and (b) whether or not a natural broadening mechanism  due to
(i) finite-temperature and/or  (ii) the coupling to phonons can 
give a reasonable explanation of the observed features of the ARPES spectra.

Calculations based on the $t-J$ model at
finite  temperature  have  been  done using the  Lanczos  method\cite{Jaklic},
quantum  Monte Carlo (QMC)\cite{Brunner}   and  recently   using the so-called
hybrid dynamical   momentum  average (HDMA)
method\cite{Cataudella}.  While the results obtained from these methods 
are quite useful, the conclusions drawn from any one of them should
be taken with some caution;
for example, the   Lanczos  method  can be applied to very small  size
lattices, and the so-called maximum-entropy technique which is utilized 
by QMC disregards the high energy peaks due to string excitations 
and other important details 
of the spectral function.  Though the recently used  HDMA method
takes into account the  electron-phonon vertex  diagrams and  produces 
the  broad lowest energy peak quite  well, the values of the coupling 
constant $\gamma$ considered is not strong  enough to justify the application 
of the  momentum average  method\cite{Berciu}. 

In this paper we extend the method introduced in Ref.~\onlinecite{KLR} and
developed in Ref.~\onlinecite{liu,liu2,strings}, at finite temperature
and we also include the role of optical phonons. In Refs.~\onlinecite{KLR,liu}
the boson degrees of freedom were treated within the so-called 
spin-wave approximation and their coupling to electron and 
hole degrees of freedom is linearized with respect to boson creation operators.
Furthermore, the self-consistent Dyson's equation for the single-hole
spectral function was solved within the so-called non-crossing
approximation (NCA) where only topologically ``planar'' diagrams are 
retained. In the present paper we  work within the same linearized
Hamiltonian and we include the linear coupling to optical phonons
as captured by the Holstein electron-phonon interaction. 
The calculations are carried out
at finite temperature by solving the Dyson's equation within the 
NCA for both diagrams which include propagation of spin-wave excitations
and diagrams which include phonon propagation. We find that together
the phonons with the inclusion of the thermal broadening at room 
temperature give rise to a broadened spectral function which 
exhibits similar characteristics to those found
in the ARPES studies. More precisely, the conclusions of 
Ref.~\onlinecite{strings} are valid without the need to 
artificially broaden the spectral function.
Furthermore, as it is well-known the leading vertex correction due to coupling 
to spin-waves is zero and other higher order vertex corrections 
give negligible contribution\cite{liu2}. In the present calculation
we include the leading (two-loop) vertex corrections due 
to the hole-phonon coupling and due to the coupling of the hole to
spin-wave excitations
and we conclude that their contribution allows us achieve the
same qualitative agreement with the experimentally determined
hole spectral function using smaller values of the electron-phonon
coupling constant.

In the following section (Sec.~\ref{formulation}) we
describe the formalism and the approach. In Sec.~\ref{results} we present our
results for the spectral function obtained by a numerical solution 
of the Dyson's equation. In Sec.~\ref{arpes}  we compare our
results with the experimentally determined spectral function.
In Sec.~\ref{vertex} we include the contribution of the vertex corrections
and in Sec.~\ref{conclusions} we present the main conclusions
drawn from the present study.

\section{Formulation}
\label{formulation}

%\subsection{The pure $t-J$ model at zero temperature}

The motion of a single hole  in a  spin-$\frac{1}{2}$  
 Heisenberg antiferromagnet\cite{RMP} in  a two-dimensional(2D) square  
lattice has been extensively studied using the two-dimensional (2D) 
$t-J$ model\cite{strings}:
\begin{eqnarray}
H  = -t\sum_{<i,j>,\sigma}(c_{i\sigma}^{\dagger}c_{j\sigma} +  h.c.) +
\nonumber\\\sum_{<i,j>}[JS_{i}^{z}S_{j}^{z}                         +
  \frac{J_{xy}}{2}(S_{i}^{+}S_{j}^{-} +S_{i}^{-}S_{j}^{+})].
\label{t-J}
\end{eqnarray}
The first term is the usual hole-hopping term which operates in
a space of singly occupied sites and the second and third form the
usual Hamiltonian of the Heisenberg antiferromagnet where we have allowed for
a possible anisotropy of the coupling between the $z$ and the perpendicular
spin components.
While this Hamiltonian has been thoroughly studied during the last 
almost two decades using many techniques, only a handful of methods 
are shown to yield accurate results in certain limits.
For the case of a single hole spectral function, one of such rather
successful techniques is the so-called self-consistent Born 
approximation (SCBA)\cite{liu}.

%\subsection{Optical Phonons}

A simple way to introduce the coupling of the hole motion to a single
optical phonon branch is by adding to the $t-J$ model an 
electron-phonon coupling term by means of the following 
Holstein term:
\begin{eqnarray}
H_{el-ph}=\Omega_0\sum_kb_k^\dagger
b_k+\frac{\gamma}{\sqrt{N}}\sum_{k,q}c_k^\dagger
c_{k-q}b_q+h.c.
\label{el-ph}
\end{eqnarray}
where $b^{\dagger}$ is the optical-phonon creation operator, 
$\Omega_0$ is  a characteristic optical phonon frequency and
$\gamma$ is the electron-phonon  coupling constant.  

Within the linear spin-wave  approximation, using the Bogoliubov 
transformation to diagonalize the Heisenberg term  and and 
by linearizing  the hopping term with respect to the spin-deviation  
operators, one finds\cite{KLR,liu} the following expression for the 
Hamiltonian given by Eq.~\ref{t-J}:
\begin{eqnarray}
H&=&E_{0}+J \sum_{\bf k} (f^{\dagger}_{\bf k} f_{\bf k}
+h^{\dagger}_{\bf k} h_{\bf k})+ \sum_{\mathbf{k}}\omega_{\mathbf{k}}
(\alpha_{\mathbf{k}}^{\dagger}\alpha_{\mathbf{k}}+
\beta_{\mathbf{k}}^{\dagger}\beta_{\mathbf{k}})\nonumber \\
&+&\sum_{k,q}h_{k}^{\dagger}f_{k-q}[g(k,q)\alpha_{q}+g(k-q,-q)
\beta_{-q}^{\dagger}]\nonumber\\ 
&+&f_{k}^{\dagger}h_{k-q}[g(k-q,-q)\alpha_{-q}^{\dagger}+g(k,q)\beta_{q}]+H.c.
\label{linearized}
\end{eqnarray}
The function $g(k,q)$ which plays the role of the hole-spin wave coupling
constant is defined in Refs.~\onlinecite{KLR,liu}.

%\subsection{Finite Temperature}

In order to calculate the effects of finite temperature we will use
the  Matsubara technique followed by analytic continuation  to the real  
frequencies to obtain the reduced Green's  function\cite{Mahan,Morales}. 
The self-consistent solution to Dyson's equation for the self energy of
the Hamiltonian given by Eq.~\ref{linearized}
is obtained by iterating the following equation with respect to
$n$:
\begin{eqnarray}
\Sigma_{(n+1)}({\bf{k}},\omega)~&=&~\sum_{\bf q}g^2({\bf k},{\bf q})\Bigl 
[N_{\bf{q}} G_n({\bf k-q},\omega+\omega_{\bf{q}})+ \nonumber \\
(1&+&N_{\bf{q}}) G_n({\bf k-q},\omega-\omega_{\bf q})+ \nonumber
\\ \int_{-\infty}^{\infty}&&\frac{d\epsilon}{\pi}n_{F}(\epsilon)
D_{0}({\bf{q}},\epsilon-\omega)  Im G_{r}({\bf{k-q}},\epsilon)
\Bigr ]. \label{self}\\
G_n({\bf k}, \omega) &=& {1 \over 
{\omega -\xi_{\bf k}-\Sigma_n({\bf k}, \omega)}}
\label{dyson}
\end{eqnarray}
 where $N_{\bf{q}}=n_B(\omega_{\bf{q}})$, with 
$n_B(\omega)=\frac{1}{e^{\beta\omega}-1}$, and
 $n_F(\xi_{\bf{q}})=\frac{1}{e^{\beta\xi_{\bf{q}}}+1}$. 
Also, $\xi_k=\epsilon_k-\mu$, where $\epsilon_k$ is the zeroth
order hole energy, which according to Eq.~\ref{linearized} is equal
to $J$, and $\mu$ is the chemical potential.
Here $\omega_q$ is the spin-wave frequency, $D_0$ is the spin-wave
propagator given as follows: 
\begin{eqnarray}
D_0({\bf q},\omega)={{2 \omega_q} \over {\omega^2-\omega_q^2}}.
\end{eqnarray}
and $G_{r}(\omega,k)$ is the retarded Green's function which is obtained 
from the Matsubara Green's function by analytic continuation:
\begin{eqnarray}
G_{r}({\bf k},\omega) = \lim_{\delta \to 0} G(\omega + i\delta,k).
\end{eqnarray}

In the case where we include the coupling to the optical phonons
via Eq.~\ref{el-ph} we need to add to the above expression for the 
self-energy three more terms which are the same
as the above and they are obtained from the above expressions 
by replacing the hole-spin-wave coupling constant $g$ by $\gamma/\sqrt{N}$,
and the spin-wave frequency by the phonon frequency $\Omega_0$.

While the first two terms of the above equation are of 
order unity, it can be shown that the last term is of order of $1/N$
because we only consider a single hole. This can be seen by considering the
following identity
\begin{equation}
N_h=\sum_p\int_{-\infty}^\infty n_F(\omega)A(p,\omega).
\end{equation}
The difference between this equation and the third term of
Eq.~\ref{self} is the presence of boson  propagator, $D_0$  
which is  not an  extensive
quantity.  Hence, the order of magnitude of the last term in Eq.~\ref{self}
is $\sim~O(N_h/N)$ and vanishes in the thermodynamic limit.
%\begin{widetext}
\begin{figure}[htp]
\begin{tabular}{cc}
\epsfig{file=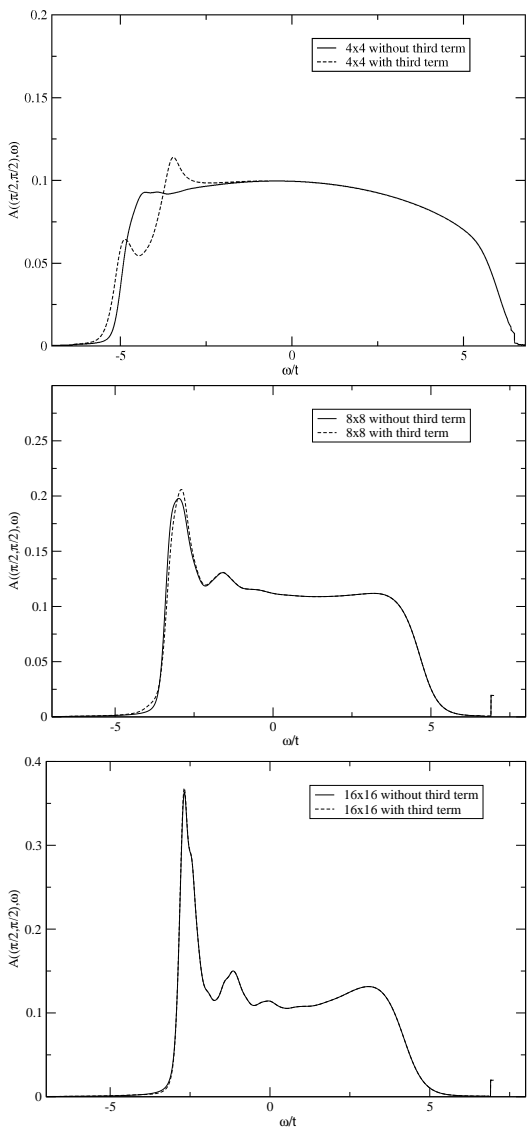,width=0.8\linewidth,clip=}\\ 
\epsfig{file=Figure1b.eps,width=0.8\linewidth,clip=}\\ 
\epsfig{file=Figure1c.eps,width=0.8 \linewidth,clip=}
\end{tabular}
\caption{The calculated spectral function for 
${\bf k} =(\frac{\pi}{2},\frac{\pi}{2})$  for  $\beta t = 10$  for a
  $4\times 4$(top), $8\times 8$(middle)  and $16\times 16$ (bottom) 
lattice with (dashed line) and without (solid line) the inclusion of the 
third term in the self-energy expression  given by Eq.~(\ref{self}).}
\label{fig1}
\end{figure}
%\end{widetext}

The vanishing of the third term in Eq.~\ref{self} is also demonstrated
numerically in Fig.~\ref{fig1} which shows the spectral function for 
$(\frac{\pi}{2},\frac{\pi}{2})$ with and
without  the third  term in Eq.~\ref{self} for  $\beta t =10$ 
and  for  $4\times 4,  8\times 8$  and $16\times 16$  lattices
respectively.  The solid lines are
the spectral functions  without the third term  while the dashed lines  
are obtained by including it by means of a single iteration
of Eqs.~\ref{self},\ref{dyson}. Notice that for large enough size lattice
the contribution of this term becomes negligible. In the rest
of our calculations presented in this paper this term
will be neglected.

The self-consistent Dyson's  equation  in conjunction with the so-called
non-crossing  approximation(NCA)
(crossing diagrams have a small  contribution as explained in 
Refs.~\onlinecite{liu,liu2}) is solved by means of an iterative procedure 
to obtain the dressed hole propagator and  
the hole spectral function.  For  numerical calculations  a  small converging
parameter $\eta$  is needed in the zeroth  order Green's
function as follows
\begin{equation}
G^{(0)}(k,\omega)=\frac{1}{\omega-\xi_k+i\eta}.
\label{zeroth}
\end{equation} 
Starting from the above zeroth order approximation for the
single hole Green's function, the Dyson's equation is iterated
until convergence is achieved. Because the lowest-energy quasiparticle peak
corresponds to a well-defined excitation, its width and height, 
as  smaller  and smaller values of $\eta$ are used,  scale
proportionally to $\eta$ and $1/\eta$ respectively. 
In addition, in order to avoid finite-size effects a smaller value
of $\eta$ requires a larger size lattice.
In Ref.~\onlinecite{liu} it was demonstrated that the single hole 
spectral function has negligible finite-size effects  for lattices larger
than  $16\times 16$  when a value for  $\eta=0.1t$ was used. 
However, when  we take  smaller values of
$\eta$, we  need bigger size  lattices to eliminate the  finite-size
effects. For example, at $T=0$ and without phonons, we have found that 
in order to reach the thermodynamic 
limit for $\eta/t =0.1$, a $16\times 16$ size-lattice is large enough, while 
if we take $\eta/t = 0.05$ or $\eta/t=0.01$ lattices of sizes $24\times 24$ 
and $32\times 32$ respectively are required.

\begin{figure}[htp]
\vskip 0.5 in
\includegraphics[width=\figwidth]{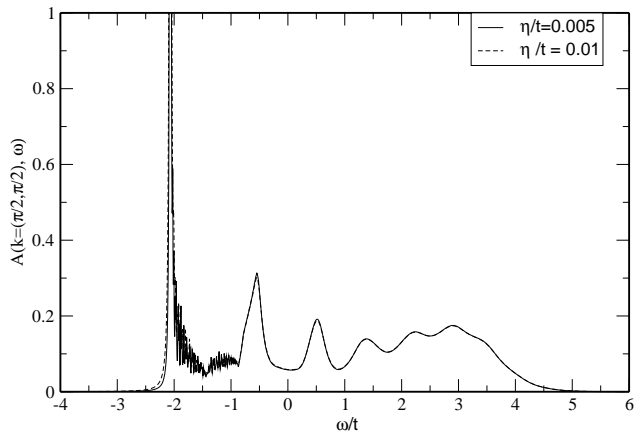}
\caption{The single-hole spectral function at $T=0$ with $\gamma=0$ for 
$(\frac{\pi}{2},\frac{\pi}{2})$ calculated on a  $64\times 64$ lattice
  with $\eta=0.005t$ (solid line) and  0.01t(dashed line).}
\label{fig2}
\end{figure}

Since the quasiparticle peak is a well-defined excitation\cite{liu} at $T=0$,
its  height becomes greater as we decrease $\eta$. On the other
hand in the  higher   energy   part    of   the   spectral function, because
it forms a continuum of states, the various Lorentzian
contributions overlap strongly because the energy spacing $\Delta \epsilon$ of
neighboring energy levels becomes exponentially small with lattice
size, i.e., $\Delta \epsilon \sim e^{-\alpha N}$ (where $N=L\times L$ is the
number of lattice sites).  Fig.\ref{fig2}     shows    the
$(\frac{\pi}{2},\frac{\pi}{2})$ spectral function with  $\eta=0.01t$ 
and $0.005t$ 
for a $64\times64$ size lattice.  The difference is mainly in the
height of  the lowest  energy peak which doubles by decreasing the 
value of $\eta/t$ by a factor of 2 and  the other  parts of both  of the
spectral functions are very close.  In this paper
we  have  used  $J=0.3t$  and  $\eta=0.05t$.

\section{Numerical Results}
\label{results}

\subsection{Finite temperature no phonons}

\begin{figure}[htp]
\vskip 0.4 in
\includegraphics[width=\figwidth]{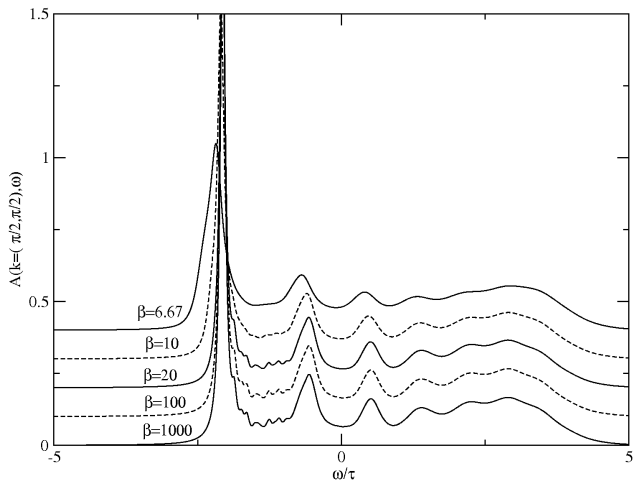}
\caption{Hole spectral function for $(\frac{\pi}{2},\frac{\pi}{2})$ without
phonons for a $16\times 16$ lattice and for $\beta t =1000$, 100,
  20, 10, and 6.67.}
\label{fig3}
\end{figure}

\begin{figure}[htp]
\vskip 0.4 in
\includegraphics[width=\figwidth]{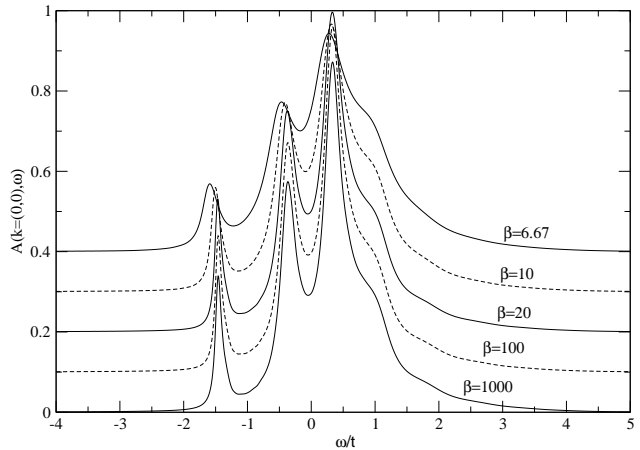}
\caption{Spectral function for ${\bf k}= (0,0)$  
without  phonons  for  a  $16\times 16$  lattice  using
  $\beta t =1000,100,20,10$ and 6.67.}
\label{fig4}
%\subfigure[ami joy]{\label{fig:edge}\includegraphics{04.eps}
\end{figure}

First of all  we study the effect of  temperature alone, i.e., without
any     phonons     in      the     system.     Fig.~\ref{fig3}     shows
the spectral function for ${\bf k}=(\frac{\pi}{2},\frac{\pi}{2})$ 
for $T/ t = 0.001,0.01,0.05,0.1$ and
0.15,  calculated on  a $16\times 16$  lattice. As  the thermal broadening
is most prominent near the lowest energy well-defined peak, notice that 
the multi-peak structure of the spectral function just  above the 
lowest peak  becomes more  and more broadened as the 
temperature is raised. The effect of  finite temperature is also to move the
low energy peaks towards lower energies (a shift of about $0.05t$ occurs
for  $T=0.1 t$).  In Fig.~\ref{fig4}, we present the spectral
 function for ${\bf k}=(0,0)$ using the same values of $\beta$. Notice that
the peaks which correspond to string excitations are robust
even for temperature as high as $T=0.15 t$ for both cases of the spectral
functions.

\begin{figure}[htp]
\vskip 0.4 in
\includegraphics[width=\figwidth]{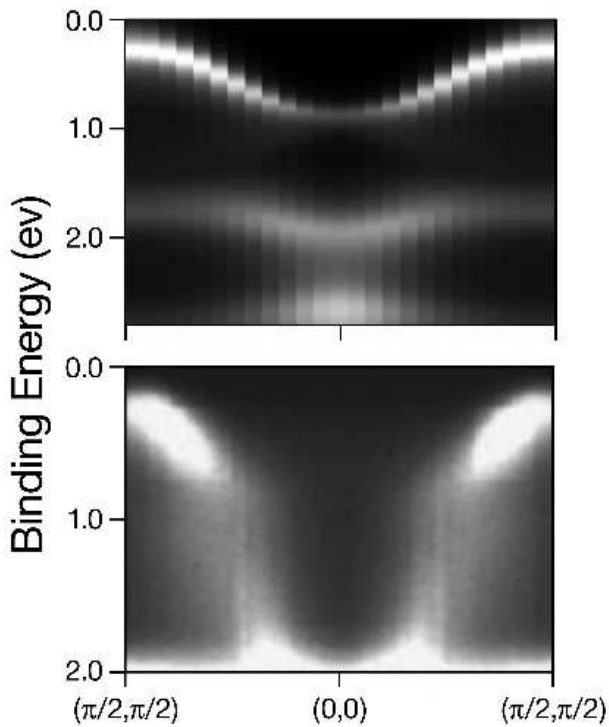}
\caption{Top: Calculated intensity  plot  on a  
$48\times 48$  lattice  at  $\beta t =10$ 
and  no  phonons. Bottom: ARPES intensity along the $(0,0)$ to $(\pi/2,\pi/2)$ 
direction. }
\label{fig5}
\end{figure}

An intensity plot for $\beta t =10$ is presented in Fig.~\ref{fig5} along
with the ARPES intensity\cite{Ronning}.
Notice that there is a significant gap or pseudo-gap between the lowest
energy peak and the first string excitation
and also  between the first string excitation and the peak which evolves
to become an intense peak near $(0,0)$. In the following we will 
discuss that the presence of 
optical phonons which couple strongly to the hole excitations
can remove these gaps and make the intensity graph similar
to those observed in ARPES.

\subsection{Optical Phonons at $T=0$}
As it has been demonstrated the NCA   is  a  good approximation  for the  
case of the single-hole in the pure $t-J$ model,  we  need   
to  make sure that it also works for the 
case of the perturbative expansion involving terms in which the
boson loops are due to the phonon propagator.  
Mishchenko  et al.\cite{mishchenko} have  shown that by  increasing 
the value of the electron-phonon  coupling strength $\gamma$  and
at zero temperature,
a cross-over between the lowest  state and the next string state takes
place at $\gamma\sim0.4t$ and from  there on the lowest state which is
like a narrow  quasiparticle peak always stays dispersionless.  
According to their calculation the next  high energy state 
shows broadening  just as it also
appears  in  the experimental  ARPES  plot\cite{Ronning}. Surprisingly  the
results of Mishchenko et al.\cite{mishchenko}
with  no  phonons  do  not  completely agree  with  earlier
numerical  results\cite{liu}.   Here,  we   will  study  the   effect  of
non-crossing  diagrams at finite temperature  using this $t-J$-Holstein
model with $\gamma$ both below and above the  cross-over point\cite{mishchenko}
and then compare our results  with experimentally obtained intensity
plots. 

Experimental values  of characteristic phonon  energy scales vary
from  $30-80  meV$\cite{damascelli,mishchenko}.  It is rather well known that
the value of $t$ is approximately $0.4eV$ for the cuprate materials 
and  we  have used
$\Omega_0\sim  0.1t$  and  $0.2t$  both producing  essentially  the 
same spectral functions\cite{mishchenko}.

\begin{figure}[htp]
\vskip 0.5 in
\includegraphics[width=\figwidth]{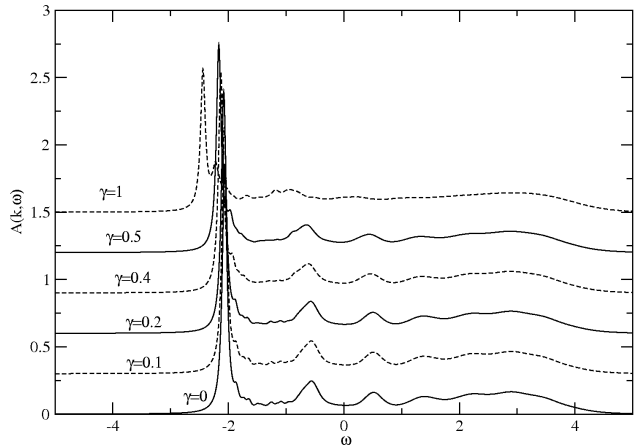}
\caption{Zero-temperature spectral
function at ${\bf k}=(\frac{\pi}{2},\frac{\pi}{2})$  on a
  $16\times 16$ size lattice   with  optical  phonons   
and $\Omega_0=0.1t$  and  for
  $\gamma$ ranging from 0 to 1.0.}
\label{fig6}
\end{figure}

\begin{figure}[htp]
\vskip 0.5 in
\includegraphics[width=\figwidth]{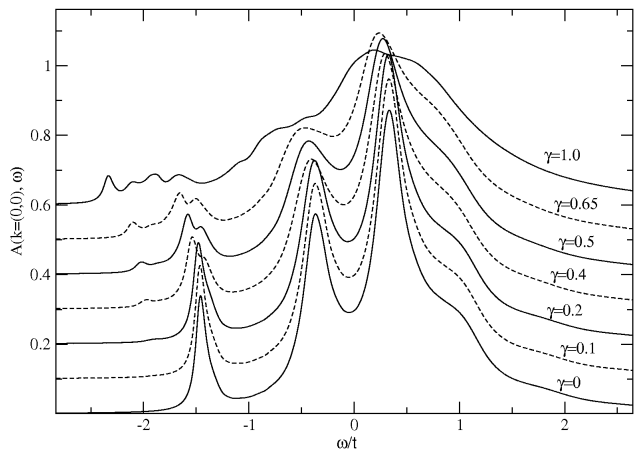}
\caption{Calculated spectral function for ${\bf k}=(0,0)$ and
at   zero  temperature for  a   $16\times 16$  size-lattice  for
  $\gamma=0.0,0.1,0.2,0.4,0.5,0.65$ and $1.0 t$.}
\label{fig7}
\end{figure}

In  Fig.~\ref{fig6} the   spectral function  for
${\bf k}=(\frac{\pi}{2},\frac{\pi}{2})$   and for
$\Omega_0=0.1t$ with $\gamma=0.0,0.1,0.2,0.4,0.5$ and 1.0 
calculated on a $16\times 16$ size lattice is presented. 
These results demonstrate that the
string excitations are quite robust in the presence of optical phonons.
In addition, in Fig.~\ref{fig7}    the   ${\bf k}=(0,0)$    spectral
function  is shown,  
for $\Omega_0=0.1t$ with $\gamma=0.0,0.1,0.2,0.4,0.5$ and 1.0
as calculated on a $16\times 16$ size lattice. The same 
conclusion about the robust nature of the string excitations 
can also be drawn from this graph.

The spectral function calculated for $J=0.3t$ has some 
saw-teeth-like features at energy just above
the  lowest peak.  Particularly  we can  see  that the  1st such  peak
closest to the lowest energy quasiparticle peak ($\sim 0.2t$ energy apart)  
gains weight as the
value of $\gamma$ is increased more  and more, a feature which is also
observed  in   DMC\cite{mishchenko,Cataudella} 
 and  in the HDMA\cite{Cataudella}   calculations   at  zero
temperature.  By increasing the value of $\gamma$ the  low energy peaks 
move towards lower energies.  For values of  $\gamma$ of about $0.5t$ 
this energy shift becomes approximately $0.1t$.  
When  $\gamma=1.0t$, however,  this energy shift becomes large  
($\sim 0.4t$).  However, NCA  is not expected to be a  
good approximation  to describe  the spectral function for
$\gamma=1.0t$. Also NCA does not show the cross-over phenomena
as  referenced by Mishchenko et al.\cite{mishchenko}.  

Phonons cannot  broaden the lowest
peak at  zero temperature (due to energy  conservation requirement)  though the
higher energy peaks corresponding to string  states are  broadened 
more and  more with  $\gamma$. In 
the next subsection (and Fig.~\ref{fig8}) we discuss that  
thermal broadening plays an important role  in filling up the
gap between the  lowest peak and the next  phonon-generated small peak 
making an
overall broad lowest energy peak.

\subsection{Finite temperature and optical phonons} 

\begin{figure}[htp]
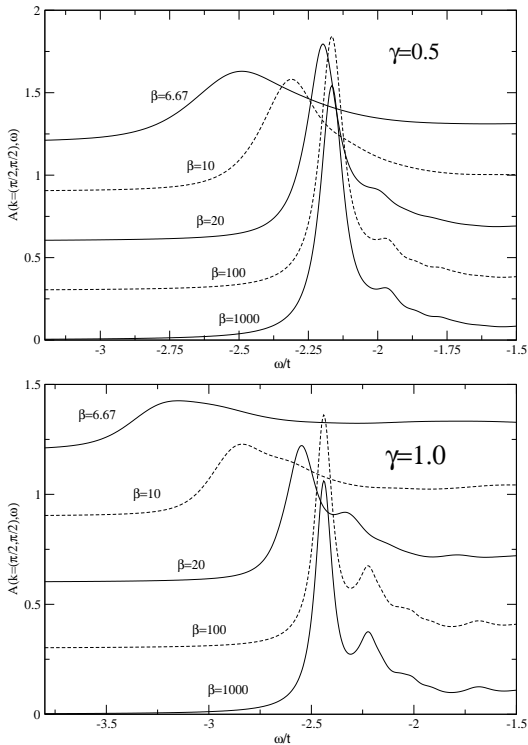

\vskip 0.5 in
\begin{tabular}{cc}
\epsfig{file=Figure8a.eps,width=.8\linewidth,clip=}\\
\epsfig{file=Figure8b.eps,width=.8\linewidth,clip=}
\end{tabular}
\caption{The spectral function for ${\bf k}=(\frac{\pi}{2},\frac{\pi}{2})$  
calculated  on a  $16\times 16$ size lattice
  for   $\beta t =1000, 100,  20,   10$   and
  6.67 at $\gamma=0.5t$(top) and $\gamma=1.0t$(bottom).
Only the vicinity of the lowest energy peak is shown.}
\label{fig8}
\end{figure}

Fig.~\ref{fig8}   shows  the vicinity of the lowest energy peak of the 
calculated spectral function for
  ${\bf k}=(\frac{\pi}{2},\frac{\pi}{2})$   for
$\beta=1000,100,20,10$ and  6.67, and   with     $\Omega_0=0.1t$    and
$\gamma=0.5t$ (Fig.~\ref{fig8}.a) and $\gamma=1.0t$ (Fig.~\ref{fig8}.b) 
 on a $16\times16$ lattice. 
Notice that for both cases of electron-phonon coupling, 
as the temperature rises the lowest peak  and the next phonon-induced peak 
smears due to thermal broadening and this gives
rise to a single broad peak as seen in Fig.~\ref{fig8}.
This is the mechanism by means of which the lowest energy peak acquires
a width of the same size as that found experimentally.
A width of similar magnitude was used in Ref.~\onlinecite{strings} to
obtain agreement with the ARPES intensity.

\begin{figure}[htp]
\vskip 0.4 in
\includegraphics[width=\figwidth]{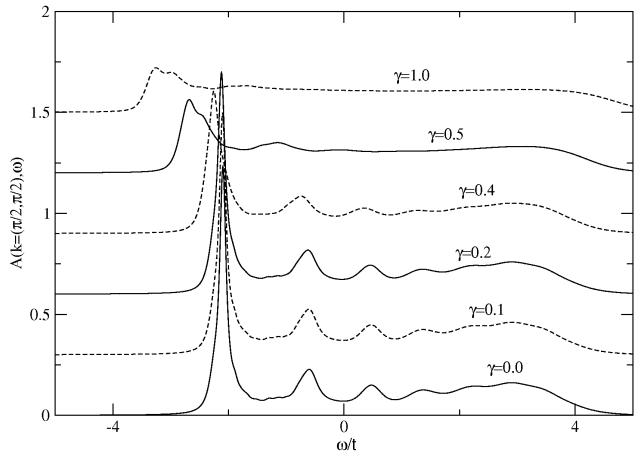}
\caption{The spectral function for ${\bf k}=(\frac{\pi}{2},\frac{\pi}{2})$  
for  $\beta=10 t$  on a  $16\times 16$  size lattice  for  $\gamma=0.0$,  
0.1,  0.2,  0.4, 0.5, and 1.0.}
\label{fig9}
\end{figure}

\begin{figure}[htp]
\vskip 0.4 in
\includegraphics[width=\figwidth]{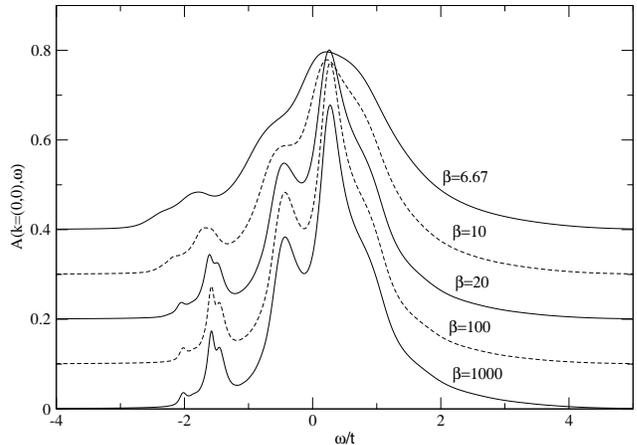}
\caption{The spectral function for $k=(0,0)$ for 
$\gamma=0.5t$  calculated on a  $16\times 16$ size lattice  and for
  $\beta t=1000,100,20,10$ and 6.67.}
\label{fig10}
\end{figure}

The effects  of the electron-phonon  interaction 
are presented in Fig.~\ref{fig9}, in a much wider frequency range, 
for ${\bf{k}}=(\frac{\pi}{2},\frac{\pi}{2})$  and for $\beta t =10$  and
 for $\gamma/t=0,0.1,0.2,0.4,0.5$ and
1.0, where   $\Omega_0=0.1t$ was used in this calculation.
In Fig.~\ref{fig10} the spectral function for $k=0$ and electron-phonon
coupling $\gamma=0.5 t$ and for various values of temperature
is shown.

\begin{figure}[htp]
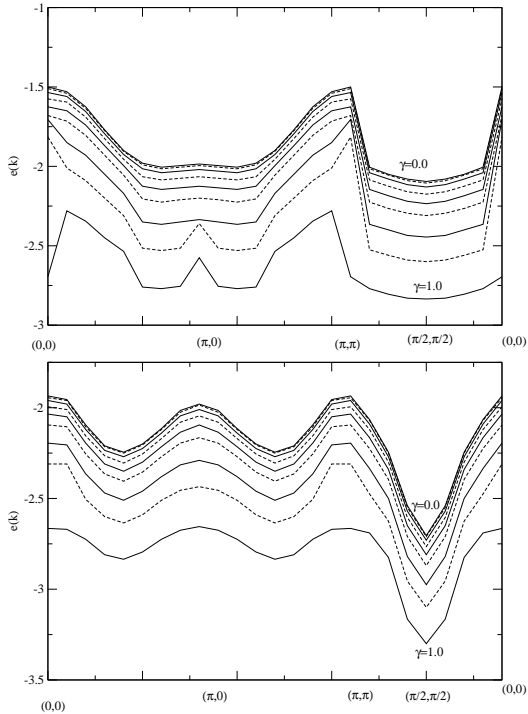

\vskip 0.4 in
\begin{tabular}{cc}
\epsfig{file=Figure11a.eps,width=0.8\linewidth,clip=}\label{c}\\
\epsfig{file=Figure11b.eps,width=0.8\linewidth,clip=}\label{d}
\end{tabular}
\caption{The hole         band  for  $\beta t=10$   and for various
values of $\gamma$ for  (top) the $t-J$
  model and (bottom) the $t-t^\prime-t^{\prime\prime}-J$ model with $t^\prime
  =-0.33t$, $t^{\prime\prime}=0.22t$. The values 
  $\gamma/t=0.0,0.1,0.2,0.3,0.4,0.5,0.65,0.8$  and 1.0 were used in 
both graphs}
\label{fig11}
\end{figure}

In Fig.~\ref{fig11}(top) we present the calculated dispersion of the 
lowest energy quasiparticle peak for the $t-J$-Holstein model
for various values of the electron-phonon coupling in the range
$\gamma/t=0-1$. In Fig.~\ref{fig11}(bottom) we present the same
calculation carried out for the $t-t^\prime-t^{\prime\prime}-J$ model
for the parameter values believed to be needed in order to reproduce
the ARPES data\cite{Kim,strings}. Notice that the only significant
effect on the $t-t^\prime-t^{\prime\prime}-J$ model, is to shift the   
overall  energy by a constant and does not alter the features of 
the dispersion. The effect of phonons on the hole dispersion for the
case of the pure $t-J$  model is more significant(Fig.~\ref{fig11}).

\section{Comparison with ARPES}
\label{arpes}

The observed ARPES spectral function reveals that the
 lowest energy  peak for ${\bf k}=(\frac{\pi}{2},\frac{\pi}{2})$
has a  width $\sim 0.4eV$ and it is the most intense feature
together with the one near ${\bf k}=(0,0)$. As we move
from $(\frac{\pi}{2},\frac{\pi}{2})$ 
towards $(0,0)$, 
the  lowest energy peak  moves gradually towards higher  energies
and a second peak grows near ${\bf k}=(0,0)$. At
around  ${\bf{k}}=(\frac{\pi}{4},\frac{\pi}{4})$ the intensity of the 
lowest energy peak decreases appreciably and  also smears over 
 a region  of higher
energy.  It is possible to explain these observations using 
the  results of the calculation based on the
$t-J$ model and the NCA as reported in Ref.~\onlinecite{strings}
where the contribution of the string excitations gives rise to
rather well-defined peaks in the spectral function at higher energy, 
provided that  these string excitation peaks broaden significantly
at around  $(\frac{\pi}{4},\frac{\pi}{4})$ to
give  rise to  some rather flat-intensity region.
Furthermore, near ${\bf k}=(0,0)$,  the peak which 
corresponds to a higher energy string excitation
suddenly picks up intensity while the broadening process
of the other  string excitation peaks still prevails. This combined process of 
spectral-weight transfer and broadening of the peaks gives rise to
the observed  energy  kinks in the ARPES intensity. 
However, these have been reported to be due to the electron-phonon  
interactions\cite{Lanzara,Cuk} and the  two  energy scales separating  the
intermediate smeared  intensity region from  the two peaks on  the two
sides (one at $(\frac{\pi}{2},\frac{\pi}{2})$  and the other at (0,0))
has  been  identified  as  the  threshold  of  disintegration  of  the
low-energy    quasi-particles into    a   spinon    and    a   holon
branch\cite{Graf}.  

\begin{figure}[htp]
\begin{tabular}{cc}
\epsfig{file=Figure12a.eps,width=0.5\linewidth,clip=}&
\epsfig{file=Figure12b.eps,width=0.5\linewidth,clip=}\\ 
\epsfig{file=Figure12c.eps,width=0.5\linewidth,clip=}&
\epsfig{file=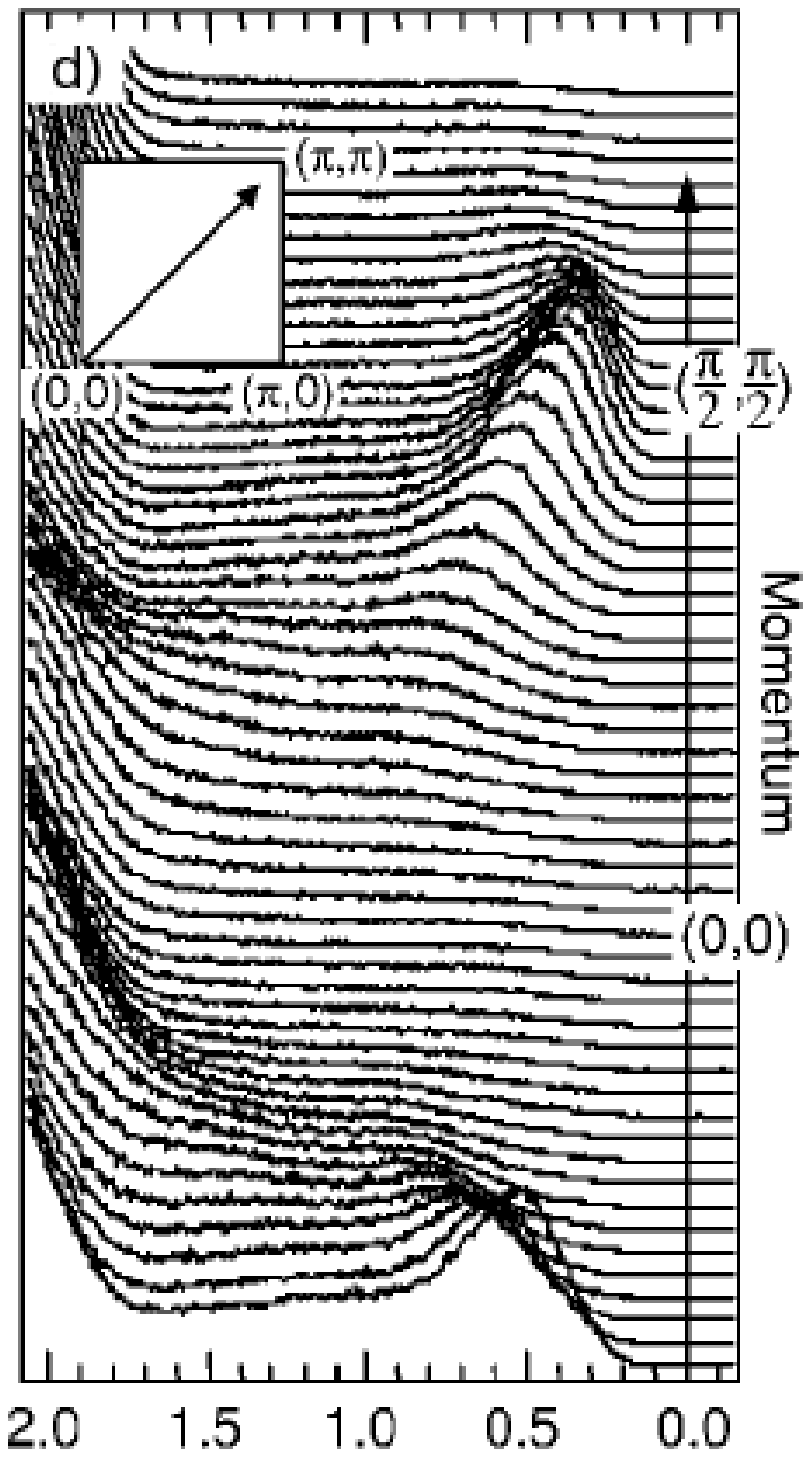,width=0.5\linewidth,clip=}
\end{tabular}
\caption{The  top-left, top-right  and bottom-left  spectral functions  
are   along (0,0)$\to(\frac{\pi}{2},\frac{\pi}{2})$  and they are calculated
for electron-phonon coupling $\gamma/t=0.2,0.5$ and 1.0 respectively
on $48\times 48$ lattice.  The  bottom-right  spectral function  is  the one  
obtained  experimentally  in  ARPES.}
\label{fig12}
\end{figure}

\begin{figure}[htp]
\epsfig{file=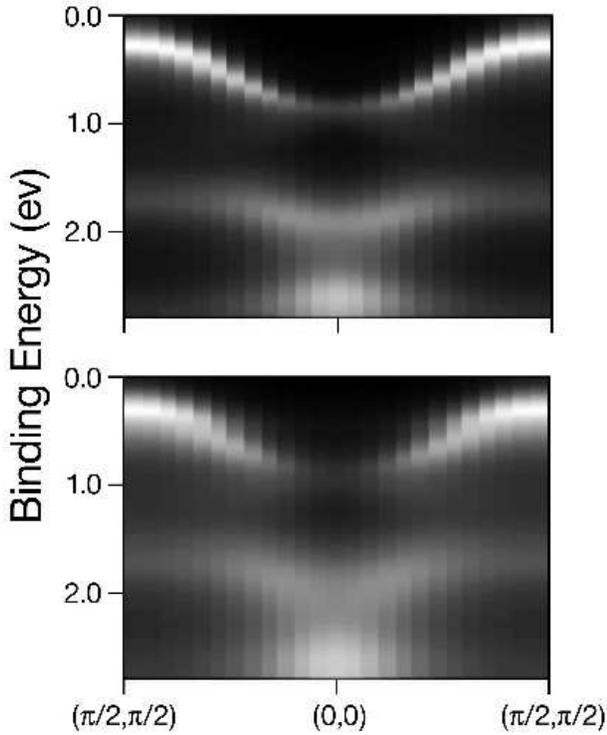,width=\figwidth}
%\end{tabular}
\caption{Intensity  plot   of  a  $48\times 48$  lattice   at  $\beta=10$  for
  $\gamma=0.2$ (top) and  0.5 (bottom)}
\label{fig13}
\end{figure}
\begin{figure}[htp]
\epsfig{file=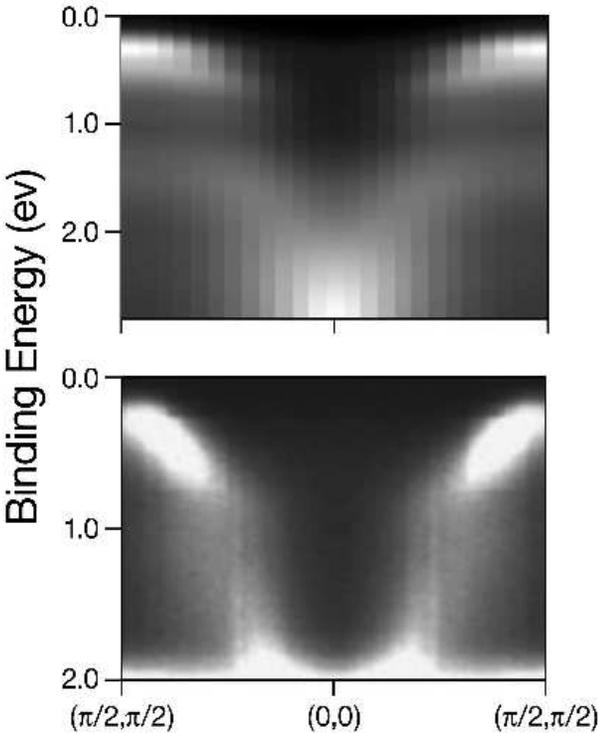,width=\figwidth}
\caption{The theoretical intensity  plot for  a  $48\times 48$  lattice   
at  $\beta=10$ and for
  $\gamma=1.0$ (top) is compared with the experimentally  obtained  ARPES
  spectra (bottom).}
\label{fig14}
\end{figure}
\begin{figure}[htp]
\epsfig{file=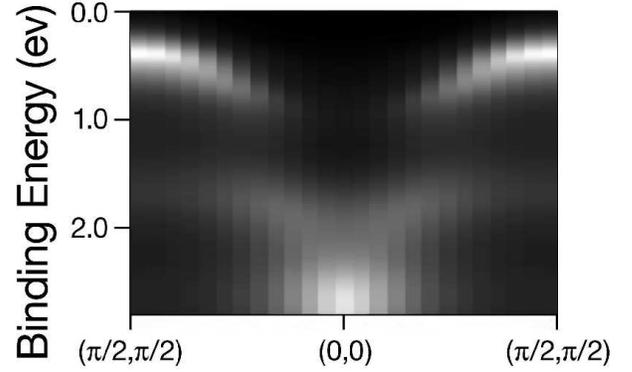,width=\figwidth}
\caption{Intensity  plot calculated on  a  $48\times 48$ lattice  
for $\gamma=0.5t$  and  $\beta t =6.67$}
\label{fig15}
\end{figure}

\begin{figure}[htp]
\epsfig{file=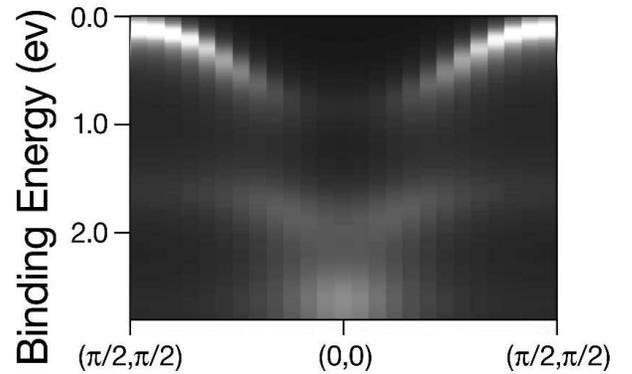,width=\figwidth}
\caption{Intensity  plot calculated on  a  $48\times 48$ lattice  
at $\gamma=0.5t$  and for $\beta t=10$
  with $t^\prime =-0.33t$, $t^{\prime\prime}=0.22t$}
\label{fig16}
\end{figure}
The  role of  temperature  is to broaden  the high  energy
string excitation peaks though the  effect is not very pronounced\cite{Kim}
without simultaneously introducing the electron-phonon coupling.   
Fig.~\ref{fig12} shows the spectral function for
all values of ${\bf k}$ along the $(0,0)\to(\frac{\pi}{2},\frac{\pi}{2})$   
direction on a $48\times 48$ lattice 
for  $\beta=10$, $\Omega_0=0.1t$ and for electron-phonon coupling
$\gamma/t=0.2$,   0.5,  and 1.0  and it is compared with experimentally   
obtained   ARPES spectral function\cite{Ronning}.
For  ${\bf{k}}$   around
$(\frac{\pi}{4},\frac{\pi}{4})$,  where the height  of the  lowest peak
and the higher energy string excitations  are almost equal, the combined 
effect of the electron-phonon  coupling and the finite temperature 
produces flat regions with much less intensity (than  that of  the lowest  
peak) in  the spectral  functions.   
Notice that using
$\gamma=0.2$ and $0.5t$,  which are below and near  the possible
cross-over point\cite{mishchenko}, the  high energy 
peak structure  due to string excitations becomes visible at around $
{\bf{k}}=(\frac{\pi}{4},\frac{\pi}{4})$
and it is broadened. This mechanism creates a  flat, low intensity region
 which becomes more pronounced for $\gamma=1.0t$.
We  note that in order for our NCA based calculation without vertex corrections
to produce spectral functions and intensity plot (Fig.~\ref{fig14}) 
similar to that obtained by ARPES we need to increase the value of  $\gamma/t$
to $1.0$  (for  $\beta t =10$, which  corresponds approximately  to
room-temperature). 
It  also   can  be  observed  that  as   we move  from
$(\frac{\pi}{2},\frac{\pi}{2})$  to  $(0,0)$  the  peak structure 
which corresponds to string excitations becomes compressed more  
and more as a function of energy and the peaks are 
closest to each other at $(0,0)$.

The  same features  can  also be  seen  in Figs.~\ref{fig13},\ref{fig14}  
which  present  the
intensity plot  on a $48\times 48$ lattice (for  $\beta=10$, $\Omega_0=0.1t$) 
and for the same three values of 
$\gamma/t=0.2$ (top part of Fig.~\ref{fig13}),  0.5 
(bottom part of Fig.~\ref{fig13}),  
 and  1.0 (top part of Fig.~\ref{fig14})   and it is 
compared with experimentally   
obtained   ARPES spectral functions (bottom part of 
Fig.~\ref{fig14}).  Notice that 
the intensity  plot becomes comparable to  the experimentally obtained
ARPES spectral function  when we use a large value of the electron-phonon
$\gamma$. As discussed in the following section, the presence
of vertex corrections due to electron-phonon-coupling allows
us to use a smaller value of $\gamma$ to achieve the same qualitative
agreement with the ARPES intensity.

Since the value of
$t$ is not accurately known, we do not have a precise knowledge of the
value of room temperature in units of $t$. 
Hence, we also tried higher temperature, $T=0.15 t$, where more
broadening is obtained.  As can be noticed from Fig.~\ref{fig15},
where $\gamma/t=0.5$ was used, temperature could be an additional
factor which helps us achieve better agreement with the observed intensity
plot without having to increase the value of $\gamma$ and enter a
domain where the validity of NCA becomes questionable.
 
As we are  mainly interested for intensity plots along
the $(0,0)$ to $(\pi/2,\pi/2)$ cut  in  the  ${\bf{k}}$-space, the inclusion  of
$t^{\prime}$  and $t^{\prime\prime}$ also  does not change  the spectral
functions very much, as  can be seen in Fig.~\ref{fig16},  
where the $t-t^{\prime}-t^{\prime\prime}-J$ model was used, taking 
$t^\prime =-0.33t$ and $t^{\prime\prime}=0.22t$ 
and in addition  $\beta=10$  and  $\gamma=0.5t$.    

\section{vertex Corrections}
\label{vertex}

The  electron-phonon  vertex  corrections  are expected to be   
important  for the  hole-spectra unlike  the electron-spin-wave  vertex
corrections    whose   contribution    have   been    found    to   be
small\cite{liu2}. There are some recent calculations indicating
the discrepancy between the spectral function  with and  without 
vertex  correction in  the strong  phonon
coupling regime at zero temperature\cite{mishchenko,Gunnarsson}.
In this section, we present the results of  our study of the role 
of such vertex corrections, namely we
improve the NCA by including the leading-order vertex corrections
due  to the electron-phonon coupling.
\begin{figure}[htp]
\vskip 0.4 in
\epsfig{file=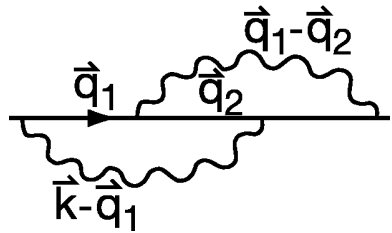,width=2. in}
\caption{The leading vertex correction to the hole Green's function
due to its coupling to phonons.}
\label{fig17}
\end{figure}
\begin{figure}[htp]
\vskip 0.4 in
\epsfig{file=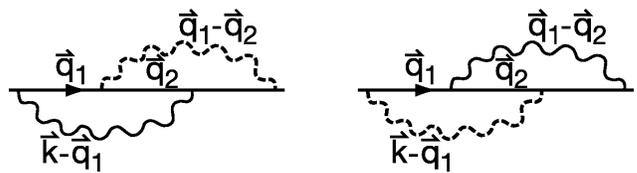,width=\figwidth}
\caption{The leading vertex corrections (two-loop) to the hole Green's function
due to phonon (solid wiggly line) and spin-wave (dashed-wiggly line)
loops.}
\label{fig18}
\end{figure}
The contributions to the self-energy are shown in the diagrams
of Figs.~\ref{fig17},\ref{fig18}
and following the procedure outlined in Ref.~\onlinecite{Mahan} we
obtain the following expressions:
\begin{widetext}
\begin{eqnarray}
\Sigma^{(\alpha)}(\omega,{\bf{k}})&=&
\sum_{\sigma=\pm 1}\sum_{\bf q_1,q_2}
G(\omega-\sigma\omega^{(p)}_{\bf{k}-\bf{q_1}},{\bf{q_1}})
\Bigl [  N^{(\alpha)}_{\bf{q_1}-\bf{q_2}}G(\omega-\sigma\omega^{(p)}_{\bf{k}-\bf{q_1}}
+\omega^{(\alpha)}_{\bf{q_1-q_2}},{\bf{q_2}})
G(\omega+\omega^{(\alpha)}_{\bf{q_1-q_2}},{\bf{k-q_1+q_2}})\nonumber\\
&+&(1+N^{(\alpha)}_{\bf{q_1}-\bf{q_2}})G(\omega-\sigma\omega^p_{\bf{k}-\bf{q_1}}-
\omega^{(\alpha)}_{\bf{q_1-q_2}},{\bf{q_2}})
G(\omega-\omega^{(\alpha)}_{\bf{q_1-q_2}},{\bf{k-q_1+q_2}}) \Bigr]
f^{(\alpha)}({\bf k}, {\bf q_1},\bf{q_2})
A^{(\sigma)}_{\bf{k}-\bf{q_1}}.  
\end{eqnarray}
\end{widetext}
The index $\alpha=1,2$ is used in order to distinguish the two 
different self-energy diagrams depicted in Figs.~\ref{fig17},\ref{fig18}.
The two different cases of $\alpha$ are obtained 
as follows:
\begin{itemize}
\item{$\alpha$=1}. For the diagram depicted in 
Fig.~\ref{fig17} which involves only phonon loops 
$\omega^{(\alpha)}_k=\omega^{(p)}_k$  where
 $\omega^{(p)}_k=\Omega_0$ is the phonon
frequency which we take it to be a constant characteristic optical
phonon frequency $\Omega_0$. In this case
$ f^{(\alpha)}({\bf k}, {\bf q_1},{\bf q_2}) =\gamma^4/N^2$. 

\item{$\alpha$=2}. For each of the diagrams depicted in Fig.~\ref{fig18} 
which involve one phonon and one spin-wave loop $\omega^{(\alpha)}_k=\omega_k$
is the spin-wave excitation frequency and 
$ f^{(\alpha)}({\bf k}, {\bf q_1},{\bf q_2}) =\gamma^2/N
g({\bf q_1},{\bf q_1-q_2})g({\bf k},{\bf q_1-q_2})$.
\end{itemize}
Here $N^{(\alpha)}_k=1/(e^{\beta \omega^{\alpha}_k}-1)$
$A^+_{\bf k}=1+N^{(1)}_{\bf k}$ and $A^-_{\bf k} = N^{(1)}_{\bf k}$
\begin{figure}[htp]
\vskip 0.4 in
\epsfig{file=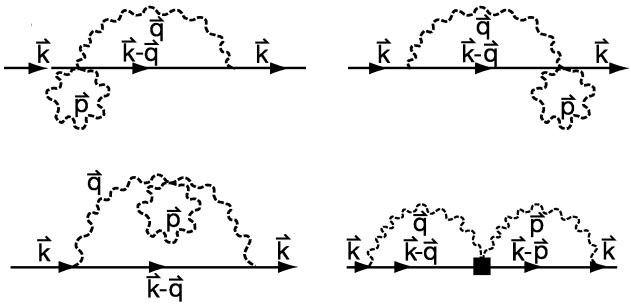,width=\figwidth}
\caption{The leading vertex corrections (two-loop) to the hole Green's function
due to spin-wave loops.}
\label{fig19}
\end{figure}
The most significant vertex corrections 
due to purely spin-wave loops are given in Ref.~\onlinecite{liu2}
and are those of Fig.~\ref{fig19}. Their contributions can be calculated
at finite temperature in a simple way as follows. The contribution of the
first two diagrams is obtained from the expression  given by Eq.~\ref{self}
by multiplying it with the prefactor $\zeta$ defined in Ref.~\onlinecite{liu2}
which is the factor that renormalizes the spin-wave velocity\cite{RMP}.
The contribution of the third diagram together with the leading order
one loop diagram (given by Eq.~\ref{self}) 
is obtained from the same expression given in Eq.~\ref{self}
by replacing the spin-wave velocity $\omega_k$ with
$(1+\zeta)\omega_k$. The last two-loop diagram of Fig.~\ref{fig19}
is given by the following expression:
\begin{eqnarray}
\Sigma({\bf{k}},\omega)&=& \sum_{\bf p,q} \rho_2({\bf q}, {\bf p}) 
 F({\bf k}, {\bf q})  F({\bf k}, {\bf p}), \nonumber \\
F({\bf k},{\bf q}) &\equiv& g({\bf k},{\bf q})  
\Bigl ( N^{(2)}_{\bf{q}} G(\omega+\omega_{\bf{q}},{\bf k-q}) \nonumber \\
&+& (1+N^{(2)}_{\bf{q}}) G(\omega-\omega_{\bf{q}},{\bf{k- q}}) \Bigr ),
\end{eqnarray}
where $\rho_2({\bf k}, {\bf q})$ is defined in Ref.~\onlinecite{liu2}.

\begin{figure}[htp]
\begin{tabular}{cc}
\epsfig{file=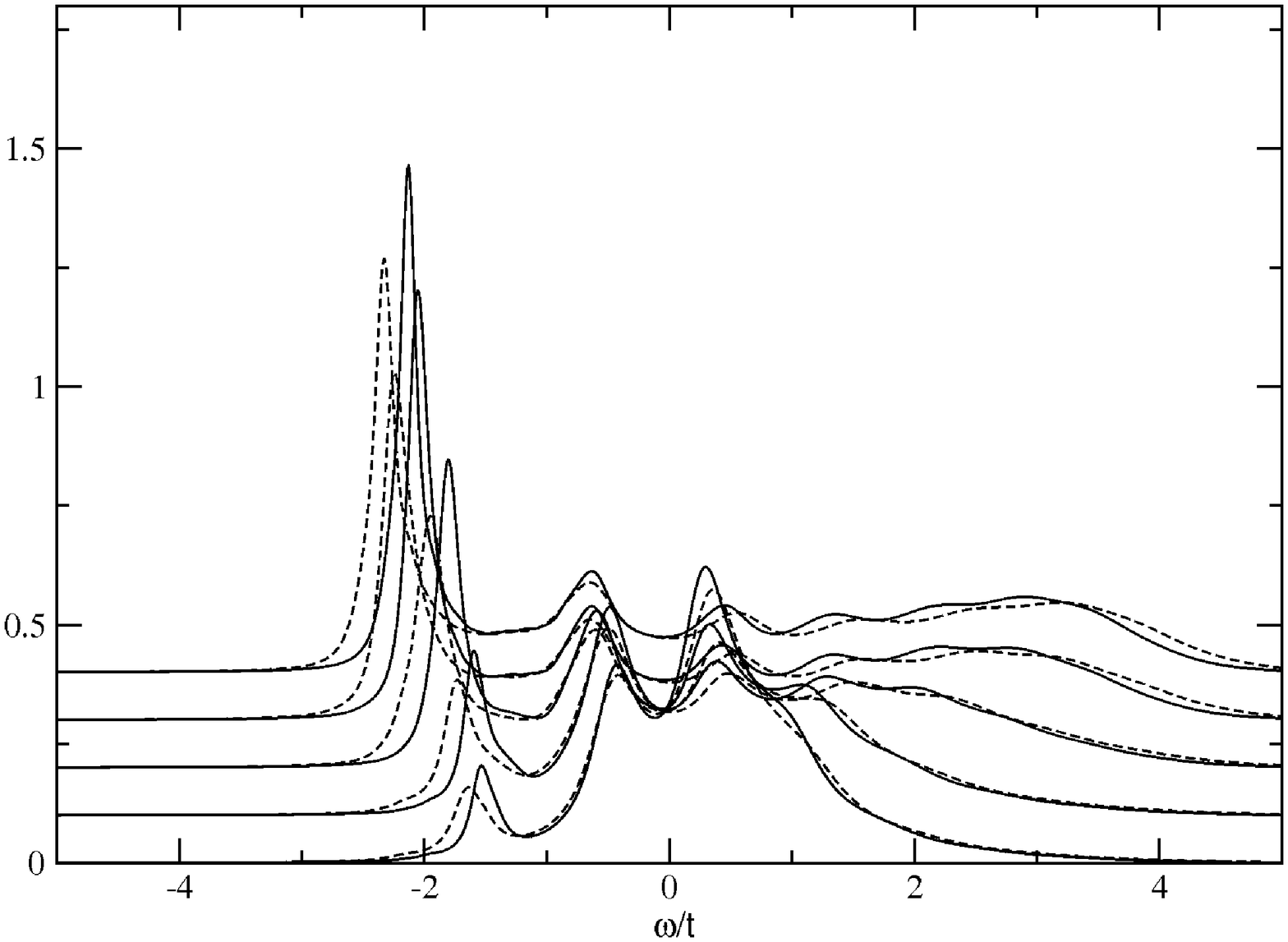,width=0.8\linewidth,clip=}\\
\epsfig{file=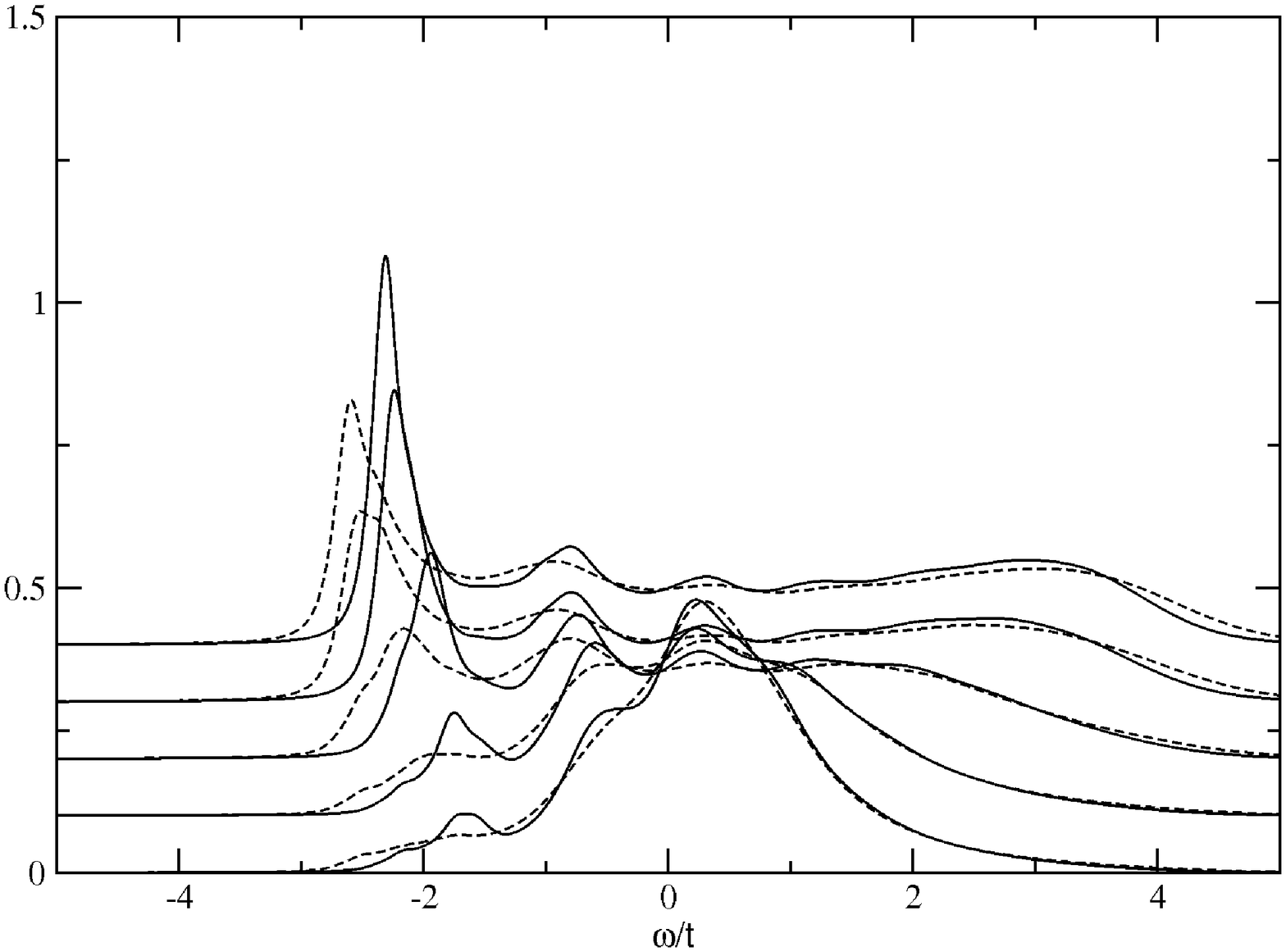,width=0.8\linewidth,clip=}
\end{tabular}
\caption{Spectral function for (a)$\gamma=0.2t$ and (b)$\gamma=0.5t$ 
for  $\beta t=10$ and on a $16\times16$ lattice with (dashed lines) and 
without (solid lines) vertex corrections due to the electron-phonon
interaction.}
\label{fig20}
\end{figure}

\begin{figure}[htp]
\epsfig{file=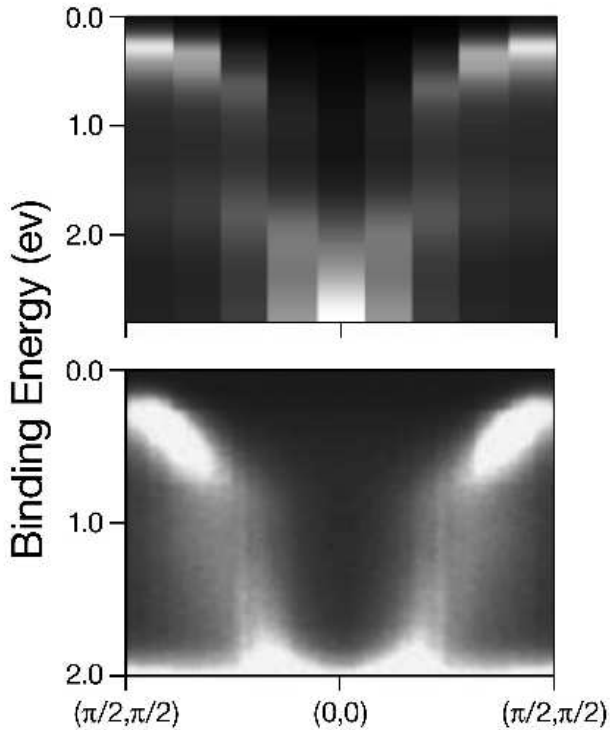,width=\figwidth}
\caption{Top: Calculated intensity  plot   of  a  $16\times 16$  lattice   
at  $\beta=10$  for
  $\gamma/t=0.5$ where vertex corrections have been included.
Bottom: ARPES intensity along the $(0,0)$ to $(\pi/2,\pi/2)$ 
direction.}
\label{fig21}
\end{figure}

We found  that  the contribution from vertex corrections to be  
small  even up to  intermediate  phonon
coupling. However,  the  difference is significant in  the strong coupling
limit.  Fig.~\ref{fig20}  shows  the  spectral function with  and  
without  the  vertex correction  using  $\gamma=0.2t$ 
(Fig.~\ref{fig20}(a))  and  0.5t (Fig.~\ref{fig20}(b)) and $\beta t=10$. 
As it can be inferred from Fig.~\ref{fig20}(b) the lowest energy peak
at  $(\frac{\pi}{2},\frac{\pi}{2})$  becomes more broadened and  
there is more rapid transfer of spectral weight as we move along the
$(\frac{\pi}{2},\frac{\pi}{2})~-~(0,0)$ direction.  Hence, owing to 
the vertex corrections the ``waterfall''-like feature, observed in
the ARPES experiments, can be reproduced
using smaller values of the electron-phonon coupling. If we include the
vertex corrections the $(\frac{\pi}{2},\frac{\pi}{2})$ peak reduces its
intensity and  at strong  enough coupling the  $(0,0)$ high  energy peak
becomes more intense than the  lowest energy peak. Therefore, we can remain in
the intermediate coupling regime (e.g., $\gamma \sim 0.5 t$ shown 
in Fig.~\ref{fig21}) and still
be able to reproduce a broadening similar to that observed in 
ARPES (bottom part of Fig.~\ref{fig21}).


\section{Conclusions}
\label{conclusions}

We have studied  the $t-J$-Holstein
model (including its $t-t^{\prime}-t^{\prime\prime}-J$ extension)  
at finite  temperature  within  non-crossing approximation.
We have also included  vertex corrections due to the 
electron-phonon coupling.  We find that the string excitations considered in 
Ref.~\onlinecite{strings} to account for the waterfall-like features
of the spectral function observed in ARPES\cite{Ronning,Graf}
are robust even at strong electron-phonon coupling and at room
temperature. Furthermore, the hole
spectral function obtained from  the NCA treatment compares well  
with the reported ARPES intensity if we adopt  a strong ($\gamma/t \sim 1$) 
hole-phonon coupling. Namely, it exhibits the same general behavior found in 
Ref.~\onlinecite{strings}, where an artificial spectral broadening was 
used in order to compare with the ARPES data; in the present treatment
this agreement is achieved without using any
such artificial broadening procedure. 
In the  calculation reported in the present paper, 
the width and the energy dispersion of the lowest energy  peak 
near $(\pi/2,\pi/2)$
is reproduced   and, in addition, we are able to
qualitatively reproduce the  high  energy  anomaly, 
i.e., the abrupt  downturn  in
intensity which is  characterised by two  energy scales
and the flat featureless intensity between them\cite{Graf}.

Our calculation, where we included the  leading-order  vertex corrections 
due to  the electron-phonon coupling,  indicates that the vertex corrections
are relatively small up to an intermediate coupling regime.
We also found that in the strong coupling limit, they become
significant, as expected. Furthermore,  in order to 
reproduce the observed features in the ARPES spectra and intensity plots 
when we included the contribution of the vertex corrections, the value 
of the electron-phonon coupling needed to achieve the same qualitative
agreement was found to be smaller than the one needed using our results 
obtained with the non-crossing approximation. 
This suggests that the qualitative features of
the results obtained within the non-crossing approximation 
(which is expected to fail in the 
strong coupling regime) might be valid in the regime describing
the cuprate materials.

\end{document}